\documentclass[sigconf]{acmart}
\usepackage{booktabs}

\renewcommand\footnotetextcopyrightpermission[1]{} 

\setcopyright{none}

\acmConference[SBES 2026]{40th Brazilian Symposium on Software Engineering}{September 8--11, 2026}{São Paulo, SP, Brazil}

\AtBeginDocument{
    
}

\usepackage{tabularx}
\usepackage{booktabs}
\usepackage{array}

\usepackage{placeins}

\newcolumntype{Y}{>{\raggedright\arraybackslash}X}

\usepackage[most]{tcolorbox}

\newtcolorbox[auto counter, number within=section]{chatquote}[2][]{
    enhanced,
    breakable,
    colback=gray!5,
    colframe=gray!55,
    boxrule=0.4pt,
    arc=2mm,
    left=2mm,
    right=2mm,
    top=1mm,
    bottom=1mm,
    before skip=4pt,
    after skip=4pt,
    fonttitle=\bfseries\footnotesize,
    title={Example~\thetcbcounter: #2},
    label={#1}
}

\begin{document}

\title{Observing the Conduct of Systematic Reviews with Generative AI Support: An Experience Report from a Graduate Software Engineering Course}

\author{Danilo Monteiro Ribeiro}
\affiliation{
  \institution{AIBL, Cesar School}
  \city{Pernambuco}
  \country{Brazil}
}
\email{dmr@cesar.school}

\author{Gilberto Sussumu Hida}
\affiliation{
  \institution{AIBL, Cesar School}
  \city{Pernambuco}
  \country{Brazil}
}
\email{gsh@cesar.school}

\renewcommand{\shortauthors}{Ribeiro and Hida}
\renewcommand{\shorttitle}{Observing the Conduct of SLRs with AI Support}

\begin{abstract}
\textbf{Context:} Secondary studies are fundamental practices in Evidence-Based Software Engineering, but teaching them requires activities that expose students to authentic methodological decisions. \textbf{Objective:} This paper reports an experience in a graduate course in which ten doctoral students in Software Engineering, organized into three groups, piloted secondary studies with and without support from generative AI. \textbf{Method:} A single-day classroom session was organized and observed, in which the groups conducted pilot systematic reviews with and without generative AI support. Classroom observations, produced artifacts, and interaction threads with assistants configured in ChatGPT were analyzed to reconstruct how each group appropriated the technology throughout the activity. \textbf{Results:} LLMs reduced initial barriers, accelerated the generation of alternatives, and made methodological problems more explicit, but they also favored excessive delegation, superficial validation, operational difficulties, and a shift in focus from conducting the SLR to using the tool. \textbf{Conclusion:} The experience offers a situated, observational account of how doctoral students engaged with generative AI during a systematic review activity, and the resulting insights also inform the design of a subsequent controlled study. The findings indicate that generative AI can support practical learning about SLRs, provided that its use is accompanied by human supervision, decision records, and critical reflection on its limitations.
\end{abstract}


\frenchspacing
\maketitle

\section{Introduction}

Systematic Literature Reviews (SLRs) and Systematic Mapping 
Studies (SMSs) are central methods in Evidence-Based Software 
Engineering (EBSE), a paradigm that advocates for grounding 
software engineering decisions in empirical 
evidence~\cite{kitchenham2004evidence,kitchenham2007guidelines}. 
These studies help organize, assess, and synthesize scientific 
knowledge in a planned, rigorous, and auditable manner, 
following a documented protocol that makes the search, 
selection, and synthesis process transparent, auditable, and 
replicable~\cite{kitchenham2007guidelines,petersen2015guidelines}. 
In graduate courses, teaching this type of study is especially 
important, since students and early career researchers need 
to understand how to select the appropriate type of review, 
formulate research questions, define search strategies, apply 
selection criteria, extract data, and synthesize evidence 
transparently~\cite{kitchenham2007guidelines,baltes2024teaching}.

This learning, however, is difficult to develop through 
theoretical exposition alone, since active and experiential 
learning tends to foster greater student engagement in 
engineering and STEM 
contexts~\cite{prince2004does,freeman2014active}. Conducting 
a review involves chained decisions, ranging from question 
formulation, protocol definition, search strategy, and study 
selection to data extraction and synthesis of 
findings~\cite{kitchenham2007guidelines,petersen2015guidelines}. 
In addition, decisions about scope, inclusion and exclusion 
criteria, and the recording of justifications are fundamental 
to the transparency, auditability, and reproducibility of the 
process~\cite{kitchenham2007guidelines,page2021prisma}. 
Problems in early stages, such as poorly formulated questions 
or inadequate search strings, may affect study selection, 
data extraction, and the structuring of the subsequent 
synthesis~\cite{kitchenham2007guidelines}. Therefore, 
practical activities are particularly useful: they allow 
students to experience the method, make mistakes, revise 
choices, and understand the consequences of the decisions 
made~\cite{kolb2014experiential,prince2004does,baltes2024teaching}.

Despite the recognized importance of systematic reviews in 
Software Engineering, Baltes and 
Ralph~\cite{baltes2024teaching} highlight that teaching 
literature reviewing remains insufficiently addressed in the 
SE education literature. Instructors face the challenge of 
designing activities that expose students to authentic 
methodological decisions, such as scope delimitation, search 
strategy construction, and evidence synthesis, without 
reducing the learning experience to mechanical protocol 
execution. At the same time, there is limited empirical 
documentation of how graduate students in SE actually engage 
with these methods when given the opportunity to practice 
them in realistic conditions. This gap is particularly 
relevant because knowing what to teach is not the same as 
understanding how students learn to conduct reviews in 
practice, including the decisions they prioritize, the 
shortcuts they take, and the points at which they struggle 
most.

The popularization of Large Language Models (LLMs) adds a new dimension to this scenario. Recent studies indicate that these models have been explored in different stages of systematic reviews, including question formulation, search strategy development, study selection, and data extraction~\cite{lieberum2025large,felizardo2024chatgpt,adam2024literature,gartlehner2024data, syriani2024screening, ferreira2026use}. In educational contexts, however, productivity gains are not the only relevant issue. It is also necessary to discuss how students interpret, validate, and incorporate model outputs, since LLM use in education involves opportunities for learning support as well as risks related to overconfidence, reduced cognitive effort, and the need for critical thinking~\cite{kasneci2023chatgpt,lee2025impact}.

This paper presents an experience report on an activity 
conducted in a graduate Software Engineering course, in 
which doctoral students were organized into three groups to 
pilot systematic reviews with and without support from 
generative AI. The focus was to observe 
how the groups appropriated the technology while performing 
real tasks of a pilot review, such as planning, question 
formulation, search string construction, search, screening, 
and initial extraction. The objective is not to causally 
measure the impact of the tool or to advocate the automation 
of reviews, but to understand how the experience was 
organized, what uses of AI emerged, what difficulties 
occurred, and what adjustments would be needed to replicate 
it in other educational contexts.

The experience revealed gains, such as reducing initial 
barriers and supporting the formulation of questions, 
strings, and criteria, but it also exposed technical 
difficulties, inconsistent responses, insufficiently 
critical acceptance, and the importance of teacher mediation 
in transforming AI use into methodological learning.

The primary audience of this report is instructors and 
researchers interested in integrating generative AI into 
Software Engineering education, particularly in graduate 
courses that cover evidence synthesis and systematic review 
methods. The contribution is not a prescription for how AI should be used, but a documented account of how doctoral students actually used it when given the opportunity, including the patterns that emerged, the risks that appeared, and the lessons that can be transferred to instructors designing similar activities. This type of situated account is particularly valuable in a moment when educators are being asked to incorporate AI tools into their practice without sufficient empirical grounding for those decisions.

\section{Theoretical Background and Context of the Experience}

The experience reported in this paper was guided by two assumptions. The first is that learning systematic reviews and systematic mapping studies in Software Engineering requires practical contact with authentic methodological decisions, such as selecting the appropriate type of review and conducting its stages according to the research context~\cite{baltes2024teaching}. These decisions include scope delimitation, research question formulation, construction of search strategies, definition of selection criteria, data extraction, and evidence analysis~\cite{kitchenham2007guidelines,petersen2015guidelines}. This understanding is aligned with the classical literature on systematic reviews in Software Engineering, which emphasizes planning, traceability, rigor, and auditability in the conduct of the study~\cite{kitchenham2007guidelines}.

The second assumption is that LLMs can be incorporated into this process as support tools, since recent studies have explored their use in different stages of systematic reviews, including search, selection, and data extraction~\cite{lieberum2025large,felizardo2024chatgpt, syriani2024screening, ferreira2026use}. However, this use needs to be critically observed, since the literature indicates that LLMs are not yet ready to replace human supervision in evidence synthesis tasks~\cite{lieberum2025large}. This caution is particularly important when students rely on the technology while they are still developing methodological expertise, since the use of generative AI in educational contexts involves opportunities to support learning, but also risks related to overconfidence, reduced cognitive effort, and the need for critical thinking~\cite{kasneci2023chatgpt,lee2025impact}. In educational contexts, recent literature on AI reinforces that the presence of technology does not eliminate the need for pedagogical guidance, response validation, and human judgment~\cite{kasneci2023chatgpt,holmes2023guidance}.

From an educational perspective, the activity was not conceived as training in a protocol for using AI in SLRs, but as a response to a contrast between manual and AI-assisted execution that students themselves had raised in earlier course sessions held without AI support. The proposal was to organize a practical situation for conducting pilot reviews, in which it would be possible to observe how groups of graduate students appropriated the technology while dealing with authentic methodological tasks. 

The common infrastructure aimed to reduce unnecessary differences between groups, supporting a consistent basis for observing how AI was used across the classroom. Even so, the way each group interacted with the technology varied throughout the activity. This variation is a central part of the report, as it reveals different modes of AI appropriation in a real educational context and helps explain how the tool was incorporated into each team's methodological decisions.

From a pedagogical perspective, AI was treated as support for the review process, and not as a substitute for methodological judgment. Even when the tool suggested questions, search terms, criteria, or screening decisions, students were expected to assess the relevance of the responses, discuss their adequacy to the review scope, and assume responsibility for the final decisions. This approach is consistent with the need for teacher mediation in activities involving emerging technology, since learning does not result only from using the tool, but from guided reflection on that use.

\section{Design and Conduct of the Activity}
\label{Design_Conduct_Intervention}

This section describes how the activity was planned and how it actually occurred in the classroom. The objective is to document the experience with a sufficient level of detail so that other educators can understand its rationale, execution conditions, and the adaptations that occurred in a real context. As previously highlighted, the focus of this section is not to compare performance between reviews with and without AI, but to reconstruct the organization of the activity and show how the interaction among students, the teacher, and LLMs materialized throughout the conduct of the pilot reviews.

It is important to clarify the nature and intent of this report. The activity was conceived as a pedagogical experience in which the alternation between manual and AI-assisted work served as a device for making methodological decisions visible to the students themselves. Our analytical interest was in the process of conduct: the quality of interactions between students and AI, the patterns of delegation, and the tensions between human judgment and model output. This observational focus is what the report documents, and it also informs the design of subsequent empirical work in this line. The 
observational richness produced by this pilot, including the 
quality of interactions between students and AI, the patterns of 
delegation, and the tensions between human judgment and 
model output, constitutes a contribution in its own right, 
independent of any experimental comparison.

\subsection{Educational and Observational Objective of the Activity}

The activity was structured around two complementary objectives. The first was to provide students with a practical experience in conducting a pilot review, allowing them to go through stages such as planning, search, screening, and initial extraction in a context close to that of a real SLR. The second was to observe how the availability of LLMs would influence the organization of group work, task distribution, artifact construction, and methodological decisions made throughout the activity.

Thus, the focus was not only on the products generated by the groups, but on the process of conducting the review. We were interested in observing when AI was invoked, what types of tasks were delegated to the tool, how students evaluated or accepted its responses, when conflicts emerged between human judgment and AI output, and how teacher mediation operated in the face of methodological doubts or operational limitations. These elements guided both the initial design of the activity and the subsequent reconstruction of what actually occurred in the classroom.

\subsection{Originally Planned Design}
\begin{figure*}[ht]
    \centering
    \includegraphics[width=0.8\textwidth]{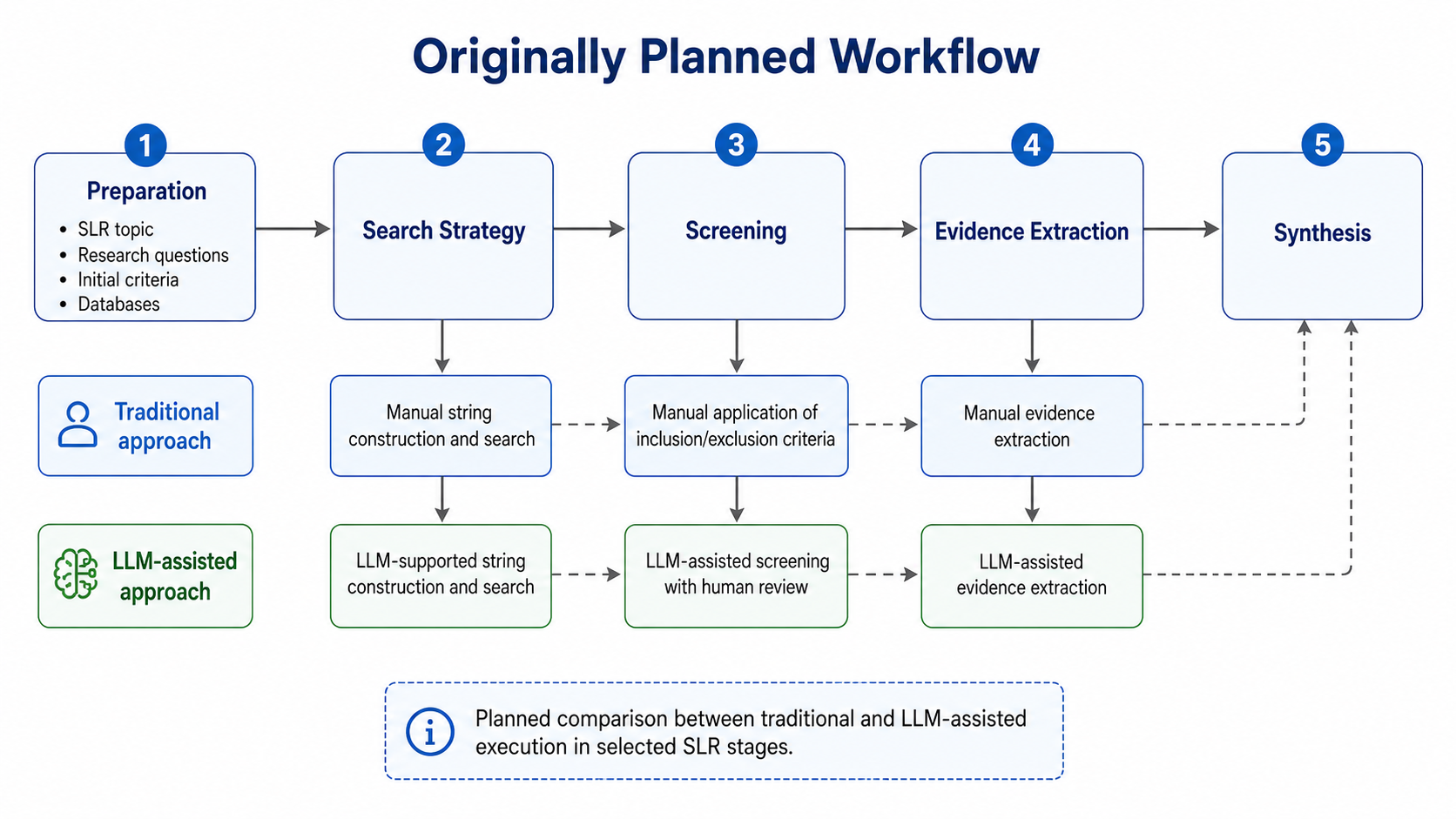}
    \Description{Flowchart illustrating the workflow originally planned for the classroom activity.}
    \caption{Originally planned workflow for the classroom activity. The design expected selected review stages to be conducted both through the traditional approach and with LLM support, allowing observation of how students interacted with each mode during the pilot reviews. Visual refinement supported by ChatGPT 5.2.}
    \label{fig:originally-planned-workflow}
\end{figure*}

The original protocol was designed around a planned comparison between manual and AI-assisted execution across the main stages of an SLR, with attention centered on how students carried out each mode rather than on comparing their outcomes. As detailed below, the analytical focus of the planned comparison would be on planning, search, screening, and extraction, without greater emphasis on the final synthesis. Figure~\ref{fig:originally-planned-workflow} presents the study workflow.

Three topics were chosen by the groups for conducting the classroom activity:

\begin{itemize}
    \item use of AI in security, focusing on the detection of attacks involving bots.
    \item use of AI in human resource management.
    \item use of generative AI in education and its ethical implications.
\end{itemize}

The original protocol structured the activity into relatively clearly delimited phases. The first phase corresponded to review planning, including definition or refinement of the topic, review objective, research questions, initial scope, and preliminary search and selection criteria. The second phase focused on the construction of search strings. In this stage, each group was expected to initially produce a minimal manual string, conduct a quick search, and then generate or refine the string with AI support, repeating the search for comparison purposes.

The third phase concerned study screening. After defining the inclusion and exclusion criteria, the groups were expected to apply the criteria manually to a subset of articles and then use AI to support the assessment of another subset or a complementary set. The protocol anticipated the recording of decisions as included, excluded, or uncertain, enabling a later contrast between human judgment and support provided by AI. Screening decisions were recorded by the groups as included, excluded, or uncertain, both when made without AI support and when supported by the tool. These records are part of the artifacts made available with this report and are used here as evidence of how the groups reasoned about selection, rather than as a basis for measuring agreement.

The fourth phase corresponded to evidence extraction. The plan envisioned an initial manual extraction in a structured spreadsheet, followed by extraction supported by AI using more structured prompts and, when possible, the upload of PDFs from the selected studies. Although synthesis was part of the complete SLR workflow, this comparative structure prioritized planning, search, screening, and extraction.

From an infrastructure perspective, in addition to the ChatGPT Plus account~\cite{openai_chatgptplus_help} with three equivalent assistants, one for each group, support materials were also made available, such as spreadsheets for search, screening, and extraction, prompt examples, guidance documents, and spaces for organizing the artifacts produced.

\subsection{Ethical Considerations}
The activity was conducted with voluntary participation, as an additional class held after the course's formal assessment had already been completed, carrying no academic grade or assessment weight. Before the activity began, all participants were informed about its research purpose and completed a consent form confirming their willingness to participate. They were also informed that they could withdraw at any moment without any consequence. Students were aware that their interactions with the AI assistants and the artifacts produced would be recorded and analyzed for research purposes. Anonymization of the data was ensured, and no individual student or group is identifiable in the findings reported in this paper.

\subsection{Practical Organization of the Classroom Activity}
\label{Practical_Organization_Classroom}
The activity was conducted in a single day, in an intensive format, with sessions in the morning (8:00–12:00) and in the afternoon (14:00–17:00). This concentrated format was adopted for logistical reasons, since several students reside in other states and a multi-session schedule would have been impractical, and it is also consistent with how courses in the program are typically organized. The activity began with the presentation of the study and the general protocol, followed by the completion of the informed consent form.

In operational terms, the groups began to work relatively autonomously, internally discussing their decisions and relying both
on the materials provided and on the LLM-based assistants. The
role of the teacher and the assistant was predominantly organizational and supervisory: clarifying general questions, maintaining the schedule, reminding students of the expected artifacts, and supporting operational issues, while avoiding direct interference in the groups' internal decisions so as not to alter the interaction dynamics intended to be observed. The activity was facilitated by two instructors with complementary backgrounds: one with experience in empirical software engineering and systematic reviews, including consulting practice on AI-assisted reviews, and another with prior teaching experience in the course and industry expertise in applied AI research. This dual profile positioned both instructors to support the methodological decisions that emerged during execution from a grounded and practical standpoint.

Throughout the class, the groups produced different types of artifacts, including refinements of research questions, search strings, preliminary lists of studies, inclusion and exclusion criteria, screening decisions, and, in some cases, initial extraction tables. In parallel, the interactions with the assistants were preserved as threads, making it possible to observe the process of artifact construction, and not only its final results, including moments in which students accepted AI suggestions, reformulated requests, faced operational limitations, sought validation, delegated decisions, or resorted to teacher mediation.

\subsection{Observed Execution}

\begin{figure*}[ht]
    \centering
    \includegraphics[width=0.8\textwidth]{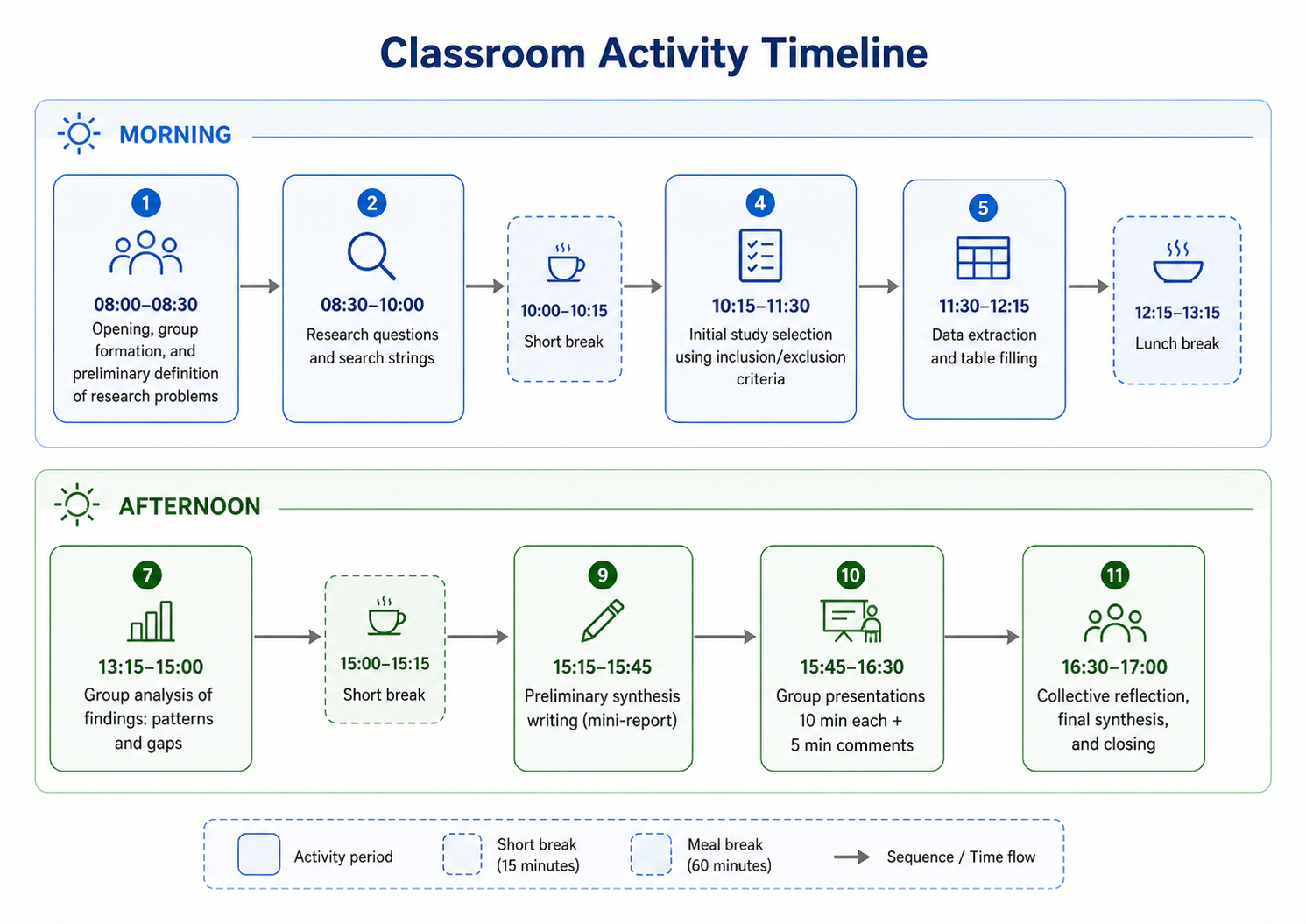}
    \Description{Timeline of the classroom activity, showing the blocks carried out throughout the day, including opening, formulation of questions and search strings, initial study selection, data extraction, analysis of findings, writing of the preliminary synthesis, group presentations, and collective reflection.}

    \caption{Timeline of the classroom activity. The figure summarizes the organization of the work blocks throughout the day, including activity periods, short breaks, and the lunch break. Visual refinement supported by ChatGPT 5.2.}
    \label{fig:activity-timeline}
\end{figure*}

The execution of the activity throughout the day can be observed in the timeline presented in Figure~\ref{fig:activity-timeline}, which summarizes the organization of the classroom work blocks, from the opening and initial formulation of the reviews to the final moments of group presentations and collective reflection. Although the groups' interaction with AI made the internal execution of each block more iterative and less linear, the teacher and the assistant kept the schedule explicit throughout the activity, issuing time alerts and guiding the transitions between the planned blocks.

\begin{figure*}[ht]
    \centering
    \includegraphics[width=0.8\textwidth]{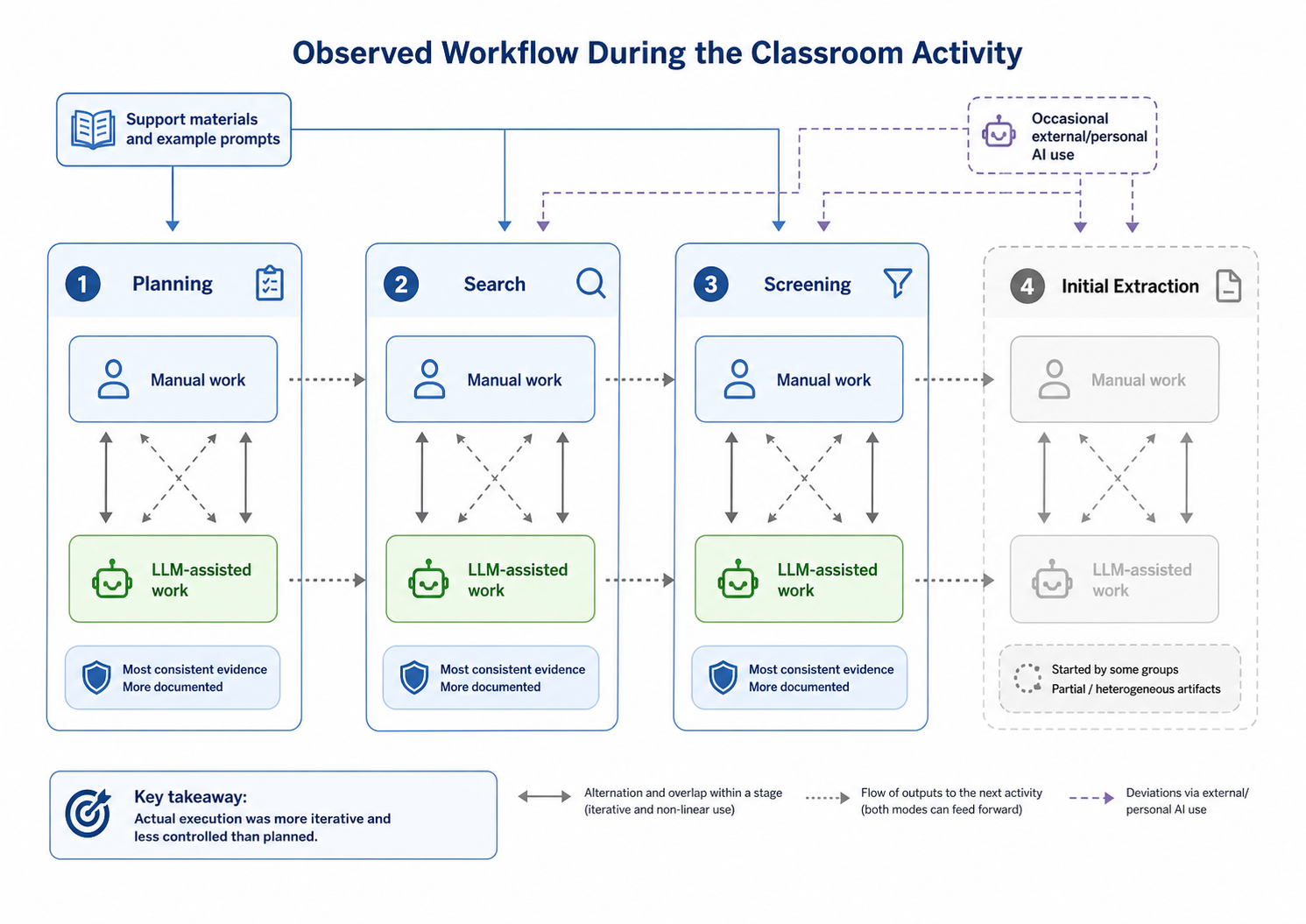}
    \Description{Diagram of the workflow effectively observed in the classroom, showing the overlap between manual work and LLM support in the planning, search, and screening stages, as well as partial initial extraction.}

    \caption{Workflow effectively observed during the classroom activity. The execution was more iterative and less controlled than planned, with overlap between manual work and LLM support; the initial extraction occurred only partially and heterogeneously. Visual refinement supported by ChatGPT 5.2.}
    \label{fig:observed-workflow}
\end{figure*}

In practice, the activity preserved the general logic of conducting pilot SLRs with AI support, but it did not fully maintain the sequence and control expected in the initial protocol. The separation between moments without AI and moments with AI proved to be less rigid than planned. Instead of always following the logic of ``manual first, then assisted'', the groups alternated or combined different forms of work throughout the activity. Figure~\ref{fig:observed-workflow} presents how the activity was conducted.

The stages that produced the most consistent data were planning, search, and screening, as illustrated in Figure~\ref{fig:observed-workflow}. In these phases, it was possible to observe more clearly how the groups formulated or refined research questions, constructed search strings, adapted queries to specific databases, defined inclusion and exclusion criteria, and made screening decisions with or without tool support. The extraction and synthesis stages, in contrast, appeared less uniformly, with greater heterogeneity among the groups and fewer artifacts generated.

At some moments, the groups started a task manually and then turned to AI to review, expand, or reorganize what had already been produced. In others, the tool was invoked from the beginning to propose central components of the review, such as questions, strings, or criteria. There were also cases in which students reused example prompts provided in the support materials, which made the boundary between the group's own methodological decision, the influence of instructional material, and the direct contribution of AI less clear.

The extraction stage was initiated by some groups, but with lower standardization. There were attempts to use AI to support article reading, structure tables, and organize evidence, but without enough uniformity to sustain a consistent comparison across groups. Synthesis, in turn, appeared in an even more embryonic form. Only one group articulated a more complete review outline, approaching a more integrated structure of planning, search, screening, extraction, and synthesis.

Situations were also observed in which AI use occurred outside the initially planned infrastructure. In some cases, operational limitations led groups to use a personal account or another AI instance, and instructors chose not to enforce a return to the institutional account once the limitation persisted, in order to preserve the naturalistic character of the observation. Since informed consent covered only the three institutional accounts, these personal-account interactions were not collected as data, a limitation we treat as a lesson learned for infrastructure planning in future replications.

\subsection{Contrast Between the Planned and the Realized}

Table~\ref{tab:planejado-realizado} summarizes the main differences between the originally planned design and the execution effectively observed in the classroom. This contrast is central to interpreting the experience, as it shows that the activity preserved the general intention of observing the conduct of pilot SLRs with and without AI support, but the separation between phases and modes of execution was less clear-cut than anticipated.

\begin{table*}[ht]
\centering
\caption{Comparative synthesis between the planned design and the execution observed in the classroom}
\label{tab:planejado-realizado}
\small
\begin{tabularx}{\textwidth}{p{3.0cm} Y Y}
\toprule
\textbf{Dimension} & \textbf{Planned in the protocol} & \textbf{Executed in the classroom} \\
\midrule

Organization of phases
&
The study was planned in relatively separate phases: planning, search, screening, extraction, and synthesis.
&
The phases did not remain fully separate. At several moments, decisions about RQs, strings, criteria, and screening occurred in an overlapping manner during the interactions with AI.
\\

\addlinespace
Manual versus AI comparison
&
The proposal was that each stage would be performed first manually and then with AI support.
&
The ``manual $\rightarrow$ AI'' sequence did not occur uniformly across groups. In some cases, the groups turned directly to AI to structure central parts of the SLR.
\\

\addlinespace
SLR planning
&
The groups were expected to define or refine the topic, objective, research questions, scope, and initial criteria.
&
All groups worked on planning elements, but with different degrees of delegation to AI. In some cases, AI played a central role in formulating or reformulating methodological decisions.
\\

\addlinespace
Construction of search strings
&
Each group was expected to construct a manual string and then a string refined with AI support, recording the retrieved results.
&
The groups used AI to construct or refine the strings. However, the existence, completeness, and comparability of the manual version varied across groups.
\\

\addlinespace
Inclusion and exclusion criteria
&
The groups were expected to define inclusion and exclusion criteria and apply them manually and with AI assistance.
&
The criteria were defined or refined with AI support to different degrees. In some groups, the final criteria were strongly influenced by the model's suggestions.
\\

\addlinespace
Study screening
&
Screening was expected to compare human judgment and judgment supported by AI, classifying studies as included, excluded, or uncertain.
&
All groups applied both screening without AI and screening with AI, but the order varied. In some cases, AI-assisted screening occurred before manual screening.
\\

\addlinespace
Evidence extraction
&
Extraction was expected to be performed manually and with AI support, using a structured spreadsheet and relating evidence to the RQs.
&
Extraction was initiated by some groups, but without sufficient standardization across all groups. AI was used to support extraction and table organization.
\\

\addlinespace
Synthesis
&
Synthesis appeared as part of the general SLR workflow, although it was not the main empirical focus of the protocol.
&
Synthesis was not consolidated uniformly. There were occasional attempts at synthesis with AI support, but only one group produced a more complete review outline.
\\

\addlinespace
Use of AI infrastructure
&
The study expected the use of one ChatGPT Plus account and three equivalently configured assistants, one for each group.
&
The planned infrastructure was used, but there were situations in which groups resorted to a personal account or another AI instance due to operational limitations.
\\

\addlinespace
Expected final product
&
The groups were expected to advance in a pilot SLR, producing artifacts across different stages of the process.
&
All groups produced relevant artifacts, but only one group produced a more complete review outline, articulating planning, search, screening, extraction, and synthesis.
\\

\addlinespace
Nature of the study effectively conducted
&
The planned design was a pedagogical structure for students to experience both manual and AI-assisted execution side by side.
&
The actual execution assumed a more exploratory character, with different patterns of AI use, delegation, and dependence across groups.
\\

\bottomrule
\end{tabularx}
\end{table*}

The first point of contrast concerns the organization of the phases. The protocol assumed a relatively linear sequence, in which planning, search, screening, extraction, and synthesis would occur as distinguishable stages. In practice, these boundaries were more diffuse: decisions about research questions, strings, criteria, and screening appeared in an overlapping manner in the groups' interactions with AI. From an educational perspective, this is relevant because it shows that conducting an SLR in the classroom does not occur merely as the application of a script, but as an iterative process of negotiation among methodological understanding, practical limitations, tool suggestions, and teacher mediation.

The second contrast involves the comparison between manual conduct and AI-assisted conduct. The planned design expected each stage to be performed first without AI and then with tool support. This sequence would allow a more controlled comparison between the two modes. However, during execution, the order varied across groups. In some cases, students turned directly to AI to structure central components of the review, such as questions, strings, or criteria. In others, they started from a manual proposal and used AI to review or expand what had been produced. This reduces the strength of a strict comparison between modes, but increases the value of the report as an observation of real appropriation of the technology in an educational context.

Variation was also observed in the degree of delegation to AI. During planning, the tool was expected to support specific tasks while preserving human judgment as the final decision-making instance. During execution, however, some groups assigned AI a more structural role, especially in formulating or reformulating questions, constructing strings, and defining criteria. This point is important because it highlights a pedagogical risk: artifacts that appear well structured may reflect the fluency of the tool or the prompt templates rather than students' methodological appropriation.

The planning, search, and screening stages produced the most consistent evidence about the interaction between students and AI. In these stages, it was possible to observe more clearly how the groups requested support from the tool, accepted suggestions, adjusted decisions, and recorded artifacts. Extraction and synthesis, in contrast, appeared in a more partial and heterogeneous way. Thus, although they are part of the general SLR workflow, these stages should be interpreted in the report as records of attempt and exploration, rather than as phases sufficiently homogeneous to support systematic comparison across groups.

Another relevant deviation concerned the use of the infrastructure: as already noted, operational limitations led some participants to use personal accounts or other AI instances, which reduces the uniformity of the conditions originally planned for the activity and reinforces a practical lesson: educational activities with LLMs need to anticipate technical contingencies, access limits, file processing issues, and alternative forms of recording.

This contrast does not aim to invalidate the experience, but to make
explicit the real conditions under which the activity occurred, and
constitutes one of the central sources of learning in this report.

\section{What Worked, What Failed, and What We Learned}

The previous sections described the context, planning, and observed execution of the activity. In this section, we shift the focus from the structure of the study to the classroom experience: how the groups used AI, what benefits emerged, what difficulties appeared, and what pedagogical tensions became visible. The objective is not to evaluate the final quality of the SLRs produced, but to reflect on what the activity revealed about the interaction among students, the teacher, support materials, and generative AI during the conduct of pilot reviews.

The analysis reported in this section draws on three sources of evidence: field notes from classroom observation, artifacts produced by the groups (research questions, search strings, selection criteria, and extraction tables), and the interaction threads preserved from the three assistants configured in ChatGPT. The threads were read in full by both authors independently, who inductively coded segments describing AI invocation, delegation, validation, and conflict between student judgment and model output, following the thematic analysis approach~\cite{braun2006thematic}. Codes were then grouped into the categories presented in Section 4.1, and discussed iteratively until consensus was reached. Illustrative excerpts were selected to represent patterns that appeared across multiple interactions, not isolated occurrences. When possible, findings were corroborated across sources: patterns identified in the threads were cross-checked against the artifacts produced and the classroom observations recorded during the activity. No formal coding software was used; the analysis was interpretive and inductive, consistent with the exploratory nature of this experience report.

\subsection{Observed Types of AI Use}

The threads indicate that AI use varied between methodological support and delegation of responsibilities. At some moments, the groups used the tool as a critical interlocutor, requesting review, suggestions, or refinements of decisions previously formulated by the students. At others, AI began to assume a more structural role, producing complete artifacts or conducting central parts of the review, such as research questions, strings, inclusion and exclusion criteria, screening, and initial extraction. This variation is important because it allows discussion not only of which tasks AI was used for, but also of who was effectively conducting the methodological decision at each moment: the students, the tool, or a poorly delimited combination of both.

A positive and expected use occurred when AI was invoked as a review or refinement assistant. In this case, students still started from an initial formulation and turned to the tool to receive comments, alternatives, or suggestions for improvement. Groups 1 and 3 illustrate this pattern by requesting assessments of the chosen topic and search string, respectively, without asking AI to conduct the entire stage, as shown in Examples~\ref{chat:g1-tema} and \ref{chat:g3-string}.

\begin{chatquote}[chat:g1-tema]{Excerpt from the thread: Group 1}
Do you have any suggestion for improving the topic?
\end{chatquote}

\begin{chatquote}[chat:g3-string]{Excerpt from the thread: Group 3}
Adapt the search string below. Create a separate search string for each of the databases listed below.
\end{chatquote}

This type of interaction is pedagogically interesting because it keeps students as the initial authors of the decision and positions AI as support for reflection. The tool can help make ambiguities explicit, suggest delimitations, and better organize the review objective, but responsibility for accepting or rejecting the suggestions remains more clearly with the group. In this pattern, AI functions as a ``peer reviewer'' or auxiliary advisor, rather than as a substitute for decision-making.

A second pattern was the use of AI as a generator of methodological alternatives. In this case, students requested proposals for questions, strings, criteria, or analysis structures, often starting from broad topics or still initial formulations that were still insufficiently refined. This use had formative value, as it quickly produced multiple options and helped the groups visualize relevant dimensions of the review. At the same time, it already shifts part of the methodological elaboration to the tool. The risk appears when the generated alternatives are accepted as the final solution, without explicit comparison with the review scope, the RQs, or the validity criteria.

A third, more critical pattern was the delegation of structural decisions. At several moments, AI did not merely review artifacts, but produced central elements of the protocol, such as complete reformulations of RQs, search strings, inclusion and exclusion criteria, and extraction fields. Group 2, for example, asked the tool to review a string for IEEE and received a response with a diagnosis of problems, expansion of synonyms, adjustment of terms, and a new version of the query. This type of support can be useful, but it also requires students to understand why certain terms were included or removed, since the string directly defines the space of retrieved studies.

Delegation became even more evident in the definition of selection criteria. In Group 1, the request made to AI explicitly indicates an intention to facilitate screening through more exclusion criteria, as shown in Example~\ref{chat:g1-criterios}.

\begin{chatquote}[chat:g1-criterios]{Excerpt from the thread: Group 1}
Create inclusion and exclusion criteria. Using just IEEE. Use more exclusion than inclusion, to make selection easier.
\end{chatquote}

\begin{chatquote}[chat:g3-extract]{Excerpt from the thread: Group 3}
Run the screening across all 164 rows and return the CSV file containing the classifications directly.
\end{chatquote}

Group 3 shows the same pattern in a more extreme form, delegating a complex extraction task to AI without specifying criteria, fields, or validation steps (Example~\ref{chat:g3-extract}). These examples are relevant because they do not represent only technical requests; they reveal a pedagogical problem. On the one hand, the students sought to make the activity feasible within the available time, which is understandable in an intensive class. On the other hand, the formulation suggests that operational convenience could guide the construction of the criteria, without an explicit discussion of coverage, selection bias, or the risk of excluding relevant studies. Thus, AI begins to support a methodological decision that should be justified by the group, and not merely optimized to reduce effort.

In summary, the observed uses indicate that AI alternated among three roles: refinement assistant, alternative generator, and decision executor. The first role was the most aligned with the pedagogical objectives, as it favored reflection and review. The second had formative potential, but required careful validation. The third revealed the greatest risk of cognitive dependence, especially when central SLR decisions were delegated to the tool without sufficient methodological justification by the students. Table~\ref{table_synthesis_ai_patterns} summarizes these three patterns, their pedagogical risk, and their associated level of criticality.

\begin{table}
\caption{Synthesis of observed AI use patterns}
\label{table_synthesis_ai_patterns}
\begin{tabular}{p{1.4cm}p{2.6cm}p{1cm}}
\toprule
Pattern & Pedagogical risk & Criticality \\
\midrule
Refinement assistant & Low, students retain authorship & Low \\
Alternative generator & Acceptance without comparison to scope or RQs & Medium \\
Decision executor & Cognitive dependence, unjustified delegation & High \\
\bottomrule
\end{tabular}
\end{table}

\subsection{Perceived Benefits and Key Learning Moments}
\label{Perceived_Benefits_Learning_Moments}

One of the most visible benefits of AI was the reduction of the initial barrier to producing artifacts. In an SLR activity, students often face difficulty moving from broad formulations, such as ``state of the art'', to more observable and auditable questions. AI helped the groups produce more structured initial versions by offering alternative wording, dimensions of analysis, and methodological justifications. This support was especially useful in the stages of planning, RQ formulation, and string construction.

Another gain was the explicit identification of methodological problems that could remain implicit in a traditional class. When AI evaluated an RQ or a string, it often returned a list of strengths, problems, and suggestions. Even when the response was not perfect, it provided material for discussion. In Group 2, for example, the tool evaluated the initial RQ and pointed out that the expression ``state of the art'' was not easily operationalized, that the scope was broad, and that observable dimensions were missing to guide screening and extraction.

\begin{chatquote}[chat:g2-beneficios]{Excerpt from the thread: Group 2}
What is the state of the art of the use of AI or LLM in people management?
\end{chatquote}

\begin{chatquote}[chat:g2-beneficios-resp]{Excerpt from the thread: LLM response}
Suggested adjustments (1--2 lines):
\begin{itemize}
    \item Replace ``state of the art'' with an auditable formulation: ``Which topics, contexts, and patterns of AI use (including LLMs) are reported in people management (2020--2025, EN)?''
    \item Include explicit dimensions of analysis: technologies, functions/levels, governance/risk, value metrics.
\end{itemize}
\end{chatquote}

AI also favored the rapid generation of alternatives. Instead of producing a single formulation of an RQ, string, or criterion, the groups could request variations, compare options, and choose paths. This made the class more active and exploratory: students did not merely receive explanations about SLRs, but tested formulations, observed responses, adjusted requests, and recorded artifacts. From a pedagogical perspective, this was one of the most positive aspects of the experience, as the tool accelerated cycles of trial and revision.

There were also benefits in the classroom dynamics. Compared with a purely expository activity, the use of LLMs made the environment more interactive. The groups began to negotiate decisions among themselves and with the tool, while the teacher monitored doubts, intervened in impasses, and drew attention to risks. In this sense, the high point of the class was not only the speed with which the groups produced artifacts, but the possibility of transforming AI responses into objects of discussion.

\subsection{Difficulties, Errors, and Human--AI Conflicts}
\label{Difficulties_Errors_and_Human}
Despite the benefits, the experience also revealed important difficulties. The first was operational. There were access problems, processing limits, difficulties with files, and the need to adapt formats. In particular, the batch screening request from Group 3 shown in Example~\ref{chat:g3-extract} exposed practical limitations of the environment: the tool stated that it could not directly open the spreadsheet and suggested exporting it or running a script locally, illustrating a typical conflict between the expectation of full automation and the actual limitations of the tool.

The second difficulty involved the literacy required to interact with AI. Although the support materials provided prompt examples, the groups still had to deal with syntax, output instructions, file formats, decision rules, and context restrictions. In some cases, the prompts became long and rigid, increasing the appearance of control over the tool, but without ensuring that it would follow exactly the expected format. This generated rework, successive adjustments, and, in certain situations, shifted the focus of the activity: instead of concentrating efforts on conducting the SLR, the groups spent significant time trying to discover how to make the LLM execute the task, process the files, or return the output in the desired format.

The third difficulty was methodological. At several moments, AI offered responses with a convincing tone and an appearance of rigor, but without the groups sufficiently discussing the implications of the decisions. The example from Group 1, shown in Example~\ref{chat:g1-criterios}, in which more exclusion criteria were requested to facilitate selection, is illustrative: the tool validated the strategy as a way to simplify screening, but the group did not record an explicit discussion about risks of coverage loss, selection bias, or impact on the validity of the review. This type of situation shows that AI can reinforce convenient decisions if students do not critically analyze all the responses provided.

Signs of insufficiently critical acceptance were also observed. In some threads, the interaction followed a sequential pattern: students requested a stage, received the response, and moved on to the next one, with little revision of previous decisions. Short expressions such as \emph{``proceed''} or requests such as that of Group 2, which asked for a list of corrections and suggestions, suggest a flow in which AI begins to conduct a large part of the methodological progression. This does not mean a total absence of reflection, but it indicates an educational risk: the activity can produce formally organized artifacts without students necessarily appropriating the methodological reasons behind them.

\begin{chatquote}[chat:g2-lista-correcoes]{Excerpt from the thread: Group 2}
Prepare a final list with the corrections and suggestions.
\end{chatquote}

Another conflict appeared in screening. By asking for simple classifications, such as \emph{yes}, \emph{no}, or \emph{uncertain}, the groups sought efficiency and standardization. However, simplifying the output reduced the visibility of the reasoning supporting each decision. In an educational task, this is a relevant tension: the more automated and compact the output is, the fewer opportunities there are to discuss why an article was included, excluded, or marked as uncertain. Thus, the operational gain may be accompanied by a loss of formative transparency.

Finally, there was a conflict between the expectation of saving time and the time spent correcting, circumventing, or reinterpreting AI responses. The tool accelerated the initial generation of artifacts, but part of this gain was consumed by problems accessing files, reformatting outputs, adjusting prompts, inconsistencies, and the need for validation. This ambiguous balance is one of the main lessons of the experience: AI can accelerate stages of a pilot SLR, but it also introduces new tasks of control, verification, and mediation.

\subsection{Teacher Mediation and Human Supervision}

As described in Section~\ref{Practical_Organization_Classroom}, teacher mediation operated mainly at an organizational level, which helped preserve the structure of the experience even as interactions between students and AI made the internal execution of each block more iterative and less linear.

From a methodological and ethical perspective, the initial guidance emphasized that generative AI should be treated as support for conducting the review, and not as the driver of the process. However, active intervention in the internal decisions of the groups was avoided, as this could alter the dynamics intended to be observed, allowing the interactions already described in Section ~\ref{Practical_Organization_Classroom} to occur naturally.

\subsection{What We Would Do Differently and Recommendations for Replication}

The experience allowed us to identify important adjustments for future applications. The first would be to reduce the scope of the activity or distribute it over more than one meeting. The attempt to cover planning, search, screening, extraction, synthesis, presentation, and reflection in a single day made the activity intense and created pressure for shortcuts. This pressure favored the use of AI as a way to accelerate stages, but it also increased the risk of excessive delegation and superficial validation. In a new edition, it would be preferable to divide the activity into at least two moments: one dedicated to planning, search, and criteria; and another dedicated to screening, extraction, reflection, and ethical discussion.

The second adjustment would be to better prepare students to deal with the technology before the beginning of the main activity. The experience showed that the lack of specific literacy in LLMs created an additional layer of difficulty. At certain moments, the groups focused more on making the tool process files, follow formats, or return structured outputs than on discussing the review itself. A short preparatory stage, with examples of limitations, input formats, context problems, file reading, and response validation, could reduce this shift in focus.

The third adjustment would be to more clearly separate the artifacts produced without AI, with AI, and after human decision. In the original design, this separation was planned, but it was not maintained uniformly during execution. In a new version, each stage should generate three records: an initial version produced by the group, a version produced or revised with AI support, and a final decision justified by the students. This format would help preserve the comparison between modes of work and, at the same time, make the process of appropriating AI suggestions more visible.

The fourth adjustment would be to review the role of prompt templates. They were useful for guiding the activity and reducing the initial barrier, but they may also have constrained part of the interactions. In some cases, the groups followed the prompt script without necessarily discussing the methodological principles embedded in it. In a new application, prompts could be presented at different levels: a minimal prompt, an intermediate prompt, and a rigorous prompt. Students would then need to justify which version they chose and adapt the prompt to their own protocol.

The fifth adjustment would be to better plan the technical infrastructure. Activities with LLMs need to anticipate access limits, file processing, context size, spreadsheet use, CSV export, and possible tool unavailability. It would also be necessary to define in advance whether the use of personal accounts or other AI instances will be allowed and, if so, how this will be recorded. This care would avoid part of the observed deviations and increase the traceability of the experience.

Finally, we would include formal moments of reflection at the end of each stage. After planning, search, screening, and extraction, each group should briefly answer: what did we do without AI? What did AI suggest? What did we accept? What did we reject? What remained uncertain? This reflective pause would help transform the use of technology into methodological learning. Without this type of interruption, there is a risk that AI will accelerate artifact production while reducing opportunities to critically discuss the decisions made.

These adjustments also function as recommendations for instructors interested in adapting the experience: reducing the scope, making decision records explicit, planning the infrastructure, and creating moments of reflection are important conditions for AI use to support methodological learning, and not merely accelerate artifact production. These recommendations were calibrated to the profile of doctoral students observed in this activity, who already had some familiarity with research methods; instructors working with less experienced audiences, such as undergraduate students, may need to adjust the scope and pacing accordingly.

\section{Conclusion}

This paper presented an experience report on a classroom activity in
which graduate students conducted pilot reviews with and without
support from generative AI. Unlike a controlled evaluation of the
tool's effectiveness, the focus was to observe the educational experience: how the activity was planned, how it actually occurred, which uses of AI emerged, what difficulties appeared, and what lessons can support instructors interested in proposing similar activities.

As summarized in Sections~\ref{Perceived_Benefits_Learning_Moments} and \ref{Difficulties_Errors_and_Human}, the experience showed that LLMs can enrich practical SLR activities, while also introducing important pedagogical risks such as insufficiently critical acceptance of responses and delegation of structural decisions. These two faces of the same interaction, gains in fluency together with exposure to dependence and superficial validation, constitute the central empirical contribution of this report.

The main lesson of the report is that integrating AI into SLR
activities requires a balance between exploratory openness and
pedagogical control. If the interaction is excessively controlled,
the opportunity to observe real patterns of use, dependence, and
validation is lost. On the other hand, if there is no minimal structure, records, and moments of reflection, the activity may become
a simple automation of tasks, without students' methodological
appropriation of the topics developed. In this sense, AI should be
treated as support for the review process, but not as the conductor
of methodological decisions.

The report also has limitations: it involved a single graduate course, a small number of participants organized into three groups with distinct topics, a single-day execution in which the separation between AI and non-AI moments was not maintained uniformly, partial and heterogeneous extraction and synthesis stages, and occasional deviations from the planned infrastructure, such as the use of personal accounts or other AI instances. For these reasons, the findings should be read as situated reflections on an educational experience, not as causal evidence about the impact of AI on the quality of SLRs.

As future work, we intend to conduct a controlled study comparing manual and LLM-assisted SLR execution across matched groups. The present pilot informed that design in concrete ways: it revealed that phase separation requires explicit checkpoints and decision records to be maintained in practice; that students need prior AI literacy training to avoid shifting focus from the review to the tool; that screening and extraction require structured output formats defined in advance to enable meaningful comparison; and that a format conducted in a single day is insufficient to observe the full SLR workflow with adequate control. The contribution of this paper is therefore not only to document what happened, but to establish the empirical ground on which a more controlled investigation can stand.

\section*{Artifact Availability}

The artifacts of this experience, including the activity protocol, assistant configuration, anonymized interaction threads, and spreadsheets, are publicly available on Zenodo at \url{https://doi.org/10.5281/zenodo.20938953}~\cite{ribeiro_hida_artifacts_zenodo}.

\section*{Acknowledgements}
The authors thank Cesar School for the institutional and financial support provided through its internal research program, which sustains the AIBL research group to which both authors are affiliated.

\section*{Use of AI Tools}
ChatGPT 5.2 was used exclusively to support language revision, translation, textual clarity improvement, and visual refinement of the workflow figures based on structures previously defined by the authors, without contributing to the study design, data collection, analysis, interpretation of results, or formulation of scientific conclusions.


\bibliographystyle{ACM-Reference-Format}
\bibliography{sample-base}

\end{document}